# Hollow-core photonic crystal fibers for Power-over-Fiber systems

Jonas H. Osório [a, *], Joao B. Rosolem [b], Fabio R. Bassan [b], Foued Amrani [c], Frédéric Gérôme [c], Fetah Benabid [c], Cristiano M. B. Cordeiro [a]

[a] *Institute of Physics "Gleb Wataghin", University of Campinas, UNICAMP, Campinas, 13083-859, Brazil*
[b] *CPqD – Research and Development Center in Telecommunications, Campinas, 13086-902, Brazil*
[c] *GPPMM Group, XLIM Institute, University of Limoges, Limoges, 87060, France*

## A B S T R A C T

Research achievements in hollow-core photonic crystal fibers technology allow ascertaining such fibers as outstanding platforms for delivering high-power laser beams. Indeed, the key property underlying the success of this family of optical fibers for high-power beam delivery is their capability of efficiently transmitting light through empty space with minimal interaction with the fiber microstructure. In this context, here we widen the framework of hollow-core fiber-based beam delivery applications by demonstrating their utilization as promising platforms for Power-over-Fiber systems. Thus, we report on the use of a tubular-lattice hollow-core fiber to deliver a watt-level continuous-wave laser beam onto a photovoltaic converter and activate a representative camera circuit. We believe that the experiments reported herein allow identifying hollow-core fibers as eligible candidates for next-generation Power-over-Fiber devices potentially able to lift the power restrictions of current solid-core fiber-based Power-over-Fiber systems.

## 1. Introduction

The stunning recent demonstrations in the hollow-core photonic crystal fiber (HCPCF) framework consolidate this special kind of optical fiber as an enabling technology for the new state-of-the-art in photonics. Indeed, although HCPCFs guiding via the photonic bandgap (PBG) effect [1] have been the first ones to appear and to provide hope on circumventing silica-core fibers limitations – regarding, for example, the transmission of wavelengths that silica is virtually opaque and power levels that solid-core fibers cannot withstand –, it was the second generation of HCPCFs, so-called inhibited-coupling (IC) HCPCFs [2], that could push the boundaries of fiber optics technology beyond what solid-core fibers can address.

Efforts by the HCPCF community have allowed attaining several breakthroughs in the last few years. For instance, one can cite the introduction of the negative curvature concept [3], a landmark result that entailed a significant reduction of the attenuation values in kagome-lattice HCPCFs and, subsequently, the demonstration of other HCPCFs architectures displaying ultralow loss from the ultraviolet to the infrared [4-6]. Indeed, the negative curvature concept unlocked a wide variety of applications ranging from the development of ultraviolet sources [7] to telecom applications [8]. Additionally, research on IC HCPCFs allowed identifying that fibers exhibiting negative curvature core contours can further minimize the spatial overlap between the optical field of the guided core modes and the silica microstructure [3]. The tiny spatial overlap between the core mode and the silica surrounding, which can amount to part-per-million levels in IC HCPCFs [9], allows this family of fibers to act as outperforming high-power laser beam delivery platforms.

In this framework, and thanks to the minor overlap between the guided optical mode and the fiber microstructure, IC HCPCFs have been used to transport and deliver both pulsed and continuous-wave high-power laser beams. For example, femtosecond laser pulses with energies of millijoules and peak powers of giga-watts, as well as kilowatt-level continuous-wave laser beams have been successfully delivered by kagome-lattice HCPCFs [10, 11]. Additionally, very recently, nested-tubes-lattice HCPCFs have been used to transport a kilowatt-power continuous-wave laser beam at 1075 nm for 1 km [12]. Such demonstrations validate the opportunity of using HCPCFs as hosts for the transmission of laser beams at power levels that solid-core fibers cannot withstand.

In a different scenario, current Power-over-Fiber (PoF) schemes – *i.e.*, systems that employ optical fibers to transport and deliver laser beams onto photovoltaic converters, so that the optical power emitted by the laser source can be converted into electrical power for activating circuits of interest [13] – completely relies on the utilization of solid-core fibers. Indeed, it inherently constrains the power figures that are practicable in existing PoF systems to the levels that silica can tolerate (standard single-mode fibers, for example, display a fiber fuse threshold of 1.5 W at 1467 nm [14]). Notably, large-core and double-clad fibers [15, 16] have been used as alternatives to partially circumvent the power limitations in PoF. Nevertheless, by following the latter approaches, the developed systems remain intrinsically restricted by the power levels that silica can bear with.

Hence, considering the great potential of HCPCFs to act as outperforming high-power beam delivery platforms, we have very recently proposed the utilization of this special kind of optical fibers as promising platforms for PoF applications and demonstrated, to our knowledge, the first PoF experiment using a HCPCF [17]. In this manuscript, by attaining better coupling conditions than in our previous work, we use a tubular-lattice HCPCF to demonstrate the delivery of a watt-level continuous-wave laser beam at 1480 nm for PoF applications. Additionally, we perform an endurance test and show that the fiber transmission remains stable during the system operation. Finally, we provide a PoF demonstration by using the optical power transmitted by the fiber to activate a representative camera circuit. We understand that the experiments reported herein allow identifying HCPCF as promising platforms for new PoF systems. Such a novel approach allows envisaging the development of next-generation HCPCF-based PoF schemes potentially able to surpass the power limitations of current PoF systems.

---
\* *Corresponding author.* jhosorio@ifi.unicamp.br

## 2. Fiber microstructure, loss characterization, and transmission endurance tests

Fig. 1a exhibits the microstructure of the fiber used in this investigation. It consists of a single-ring tubular-lattice (SR-TL) HCPCF. The fiber cladding is formed by a set of eight untouching silica capillaries with a thickness of 1050 nm and a diameter of 16 μm. The tubes in the fiber cladding define a hollow core with a diameter of 35 μm. Guidance of light through SR-TL HCPCF occurs thanks to a minimal spatial overlap between the core and cladding modes field distributions and to a strong mismatch between their transverse phases, which, in turn, entails a robust minimization of the coupling between them [18]. At specific wavelength intervals, however, due to resonances between the core and cladding modes, the fiber displays high loss. It causes the fiber transmission spectrum to alternate low- and high-attenuation regions whose spectral positions are primarily dependent on the thickness and the refractive index of the cladding tubes, and on the refractive index of the fiber core [18].

Fig. 1b presents the transmission spectra of 100 m and 4 m-long fiber pieces between the wavelengths of 700 nm and 1700 nm. The spectra in Fig. 1b have been measured for the same light coupling conditions and evaluated consecutively by following a cutback procedure. Additionally, a typical near-field profile measured at the fiber output is shown as an inset in Fig. 1b (the fiber microstructure has been placed underneath the near-field profile image so to provide better visualization of it within the fiber cross-section). Observation of the graph in Fig. 1b allows identifying two transmission bands within the 700 nm to 1700 nm wavelength range, namely between 770 nm and 940 nm and 1150 nm and 1590 nm. The loss spectrum of the fiber (calculated from the transmission traces in Fig. 1b) is shown in Fig. 1c. Within the two transmission bands, the minimum loss values are 6.0 dB/km at 830 nm and 22.4 dB/km at 1340 nm. At 1480 nm, the wavelength of the laser source used in the experiments to be described in the following sections, the fiber loss is 35.3 dB/km.

A salient feature of the light guidance in IC HCPCF is, as previously noted, the tiny overlap between the core mode and the silica microstructure. Such overlap is typically assessed via the dielectric overlap parameter ($DO$), defined by Eq. (1), where $p_z$ is the longitudinal component of the Poynting vector of the fundamental core mode, $S_{CS}$ is the fiber cross-section surface and $S_D$ is the surface within the fiber cross-section occupied by the dielectric [19].

$$DO = \frac{\int_{S_D} p_z \, dS}{\int_{S_{CS}} p_z \, dS} \qquad (1)$$

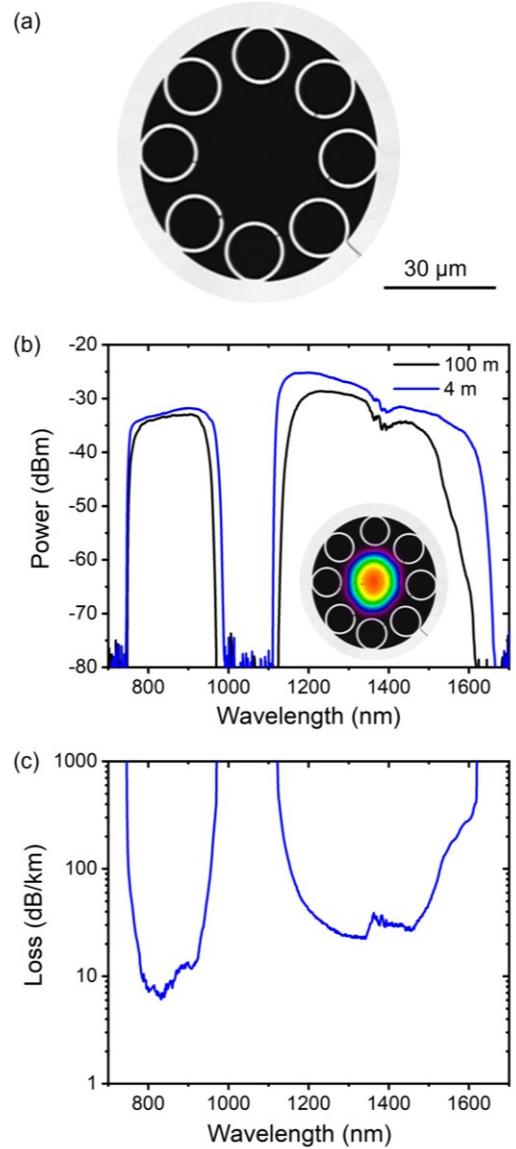

Fig. 1 (a) HCPCF cross-section. (b) Transmission spectra for fibers with lengths of 100 m and 4 m measured consecutively by following a cutback procedure. Inset shows a typical near-field profile measured at the fiber output; the fiber microstructure has been placed underneath the intensity profile for better visualization. (c) Fiber loss accounted from the cutback measurement.

The behavior of $DO$ as a function of the wavelength is such that its value increases as one approaches the edges of the transmission bands and that a minimum value of it is found when one considers wavelengths apart from the transmission cutoffs. Thus, by numerically assessing the $DO$ in representative SR-TL HCPCF, Vincetti and coworkers could determine an empirical scaling law for the minimum values of such parameter within the fiber transmission band, $DO_{min}$, as shown in Eq. (2), where $N$ is the number of tubes in the fiber cladding, $\lambda$ is the wavelength, $R_{co}$ is the radius of the fiber core, and $t$ is the thickness of the cladding tubes. By using (2) and the geometrical parameters of the fiber used in this investigation ($N = 8$, $R_{co} = 17.5$ μm, $t = 1050$ nm), one can obtain the plot shown in Fig. 2. It allows identifying that, for the fiber reported herein, $DO_{min}$ at wavelengths around 1340 nm (the minimum-loss wavelength within the transmission band between 1150 nm and 1590 nm) is expected to be at the order of $10^{-5}$ (purple rhomb in Fig. 2), *i.e.*, the power fraction of the core mode in the silica microstructure barely amounts to 0.001%. In Fig. 2, the crosshatched regions identify the transmission cutoffs.

$$DO_{min} = \left(\frac{0.24}{N}\right)\left(\frac{\lambda}{R_{co}}\right)^2 \left(\frac{t}{R_{co}}\right)^{0.93} \quad (2)$$

To confirm the power handling capabilities of the fiber used herein, one proceeded with a transmission endurance test. Thus, light from a 3W continuous-wave laser emitting at 1480 nm (IPG PYL-31480D) has been launched into a HCPCF with a length of 6 m and the transmitted power has been measured as a function of the time. The experimental setup used for the transmission endurance test is schematically represented in Fig. 3a, where L1 and L2 stand for collimation and coupling lenses, respectively, M1 and M2 represent mirrors used for alignment, and PM refers to a power meter (Newport 1835C with detector 818T-30). Fig. 3b exhibits a picture of the laser beam alignment and coupling sites (the region represented in the dashed rectangle in Fig. 3a).

Fig. 4 presents the power values at the fiber output measured for 65 minutes. During the experiments, the fiber has been kept static and coiled at a 15 cm radius. Observation of Fig. 4 allows identifying that the HCPCF afforded stable power delivery during the experiment. To better visualize the data in Fig. 4, one shows as an inset a histogram for the output power values together with a gaussian distribution curve, centered at 1.31 W and with a standard deviation of 0.01 W, which has been fitted to the experimental data. Results in Fig. 4 assure that the HCPCF can provide stable power handling at its output and, hence, that this class of optical fibers represents a promising platform for PoF applications. Additionally, it is worth saying that moderate movement of the fiber did not upset its

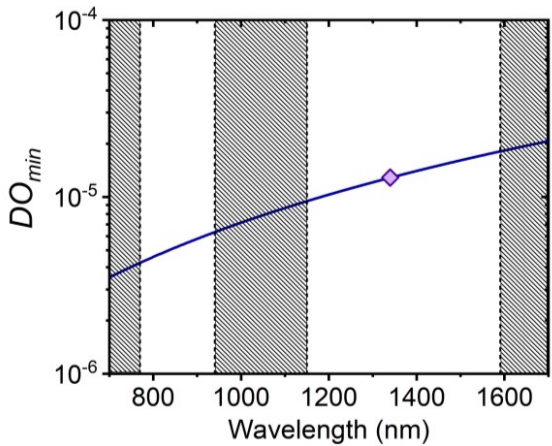

Fig. 2 Dielectric overlap parameter minima, $DO_{min}$, as a function of the wavelength, calculated from Eq. (2) by setting $N = 8$, $R_{co} = 17.5$ µm, and $t = 1050$ nm. The purple rhomb has been added for eye-guidance on the $DO_{min}$ value at 1340 nm, the minimum-loss wavelength in the transmission band between 1150 nm and 1590 nm, in which the working wavelength during the PoF experiments, 1480 nm, sits. The crosshatched regions identify the transmission cutoffs.

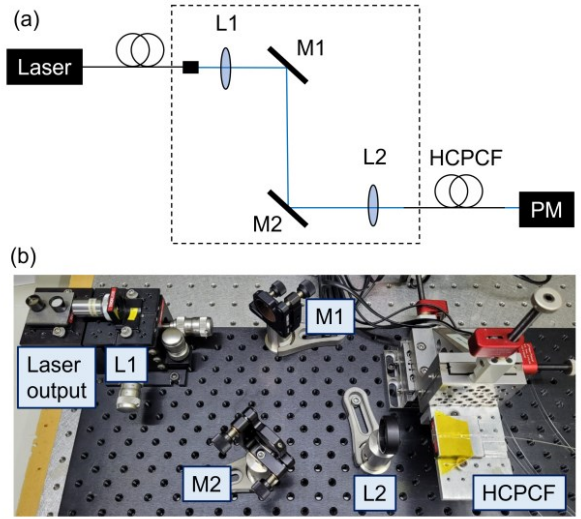

Fig. 3 (a) Schematic of the experimental setup used for the fiber transmission characterization (L1 and L2: lenses; M1 and M2: mirrors; PM: power meter). (b) Picture of the setup for light launching into the HCPCF.

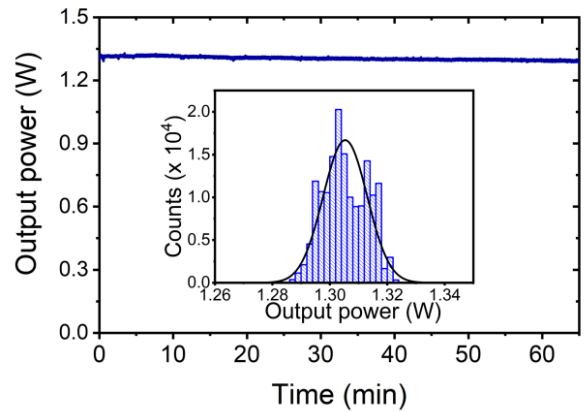

Fig. 4 Transmission endurance test results and histogram for the output power values accounted from the transmission endurance test data.

output power handling capabilities. It allows us to recognize HCPCFs as valid platforms for on-the-field operations.

## 3. Demonstration of HCPCFs as platforms for PoF applications

To illustrate that HCPCFs can successfully act as platforms for PoF applications, we carried out a representative experiment in which the laser beam transported and delivered by the HCPCF has been used to activate a micro camera circuit via the utilization of a photovoltaic converter, PV. Prior to performing the PoF demonstration, one has characterized the PV (JDSU PPC-9LW, a photovoltaic power converter optimized for maximum efficiency in the wavelength range between 1300 nm and 1550 nm) in high-power regime and obtained its corresponding electric power versus load resistance plot for different incident optical powers. The tested high-power regime was set to 7.5-fold the PV datasheet specification. Thus, the PV has been attached to a heatsink to avoid overheating.

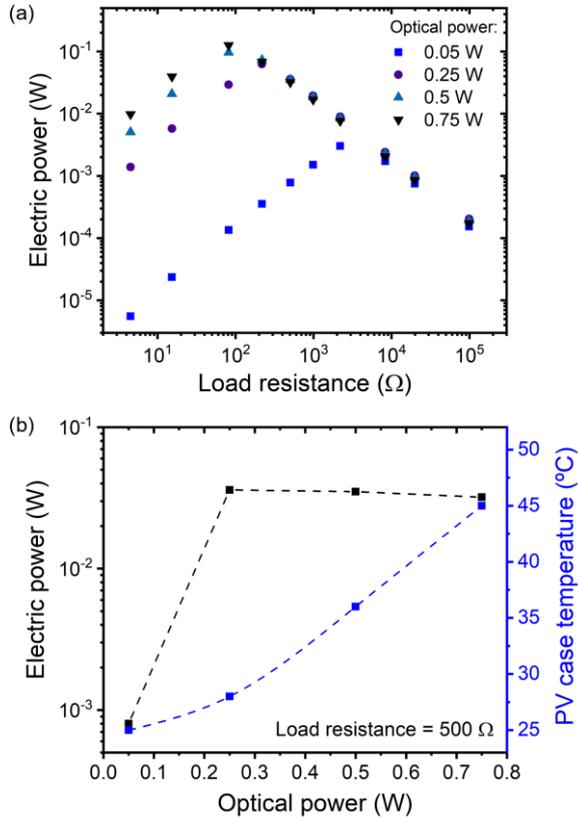

Fig. 5 (a) Graph for the electric power as a function of the load resistance connected to the photovoltaic converter terminals for different incident optical power values. (b) PV electric power and case temperature as a function of the incident optical power for a load resistance of 500 Ω.

The results of the PV characterization experiments are presented in Fig. 5a and Fig. 5b. For a given light illumination power level, maximum electrical power is attained for a particular combination of electric current and voltage values so that there is an optimum load resistance achieving maximum power transfer from the PV to the load [20]. Results in Fig. 5a identify a typical PV response. Since the electrical power is obtained by multiplying the voltage and electric current values, its maximum value is found for an optimum load resistance larger than zero (short-circuit condition) for which the corresponding voltage is close to the open-circuit voltage. In this context, data in Fig. 5a shows that, for example, for a 0.75 W incident optical power, the PV provides a maximum electric power level of 0.125 W for an optimum load resistance of 82 Ω. For load resistances larger than the optimum, the voltage level remains close to the open-circuit voltage. However, the increase of the load resistance causes the current intensity to decrease, which entails, in turn, a reduction in the electric power levels. Thus, as the different PoF applications one might consider may encompass devices with characteristics that do not fulfill the conditions for attaining maximum power, it is crucial to verify the PV characterization curves to determine whether the PV device can afford the necessary power levels for the desired application.

Additionally, Fig. 5b shows the PV electric power and the PV case temperature (measured by using a thermopile placed in the PV case) versus the incident optical power for the situation in which a load resistance of 500 Ω has been connected to the PV. This situation stands for the load resistance of the micro camera used in the experiment to be described in the following. It is seen that, for this configuration, the PV electric power grows from $8 \times 10^{-4}$ W for low input optical power (0.05 W) to a practically stable level around 0.035 W for input optical powers larger than 0.25 W (indeed, there is a slightly gradual reduction of the electric power level when the input power increases from 0.25 W to 0.75 W). This behavior is attributed to the logarithmic dependence of the PV voltage on the PV electric current and, consequently, on the input optical power [21]. The PV case temperature, in turn, increased from 25 °C to 45 °C in the studied power range. The PV case temperature of 45 °C is acceptable for the PoF demonstration described in this manuscript.

Fig. 6a shows a diagram of the experimental setup used in the PoF demonstration. The setup is similar to the one displayed in Fig. 3a (i.e., same laser source and alignment method), with the difference that, in the scheme shown in Fig. 6a, the power meter has been replaced by a PV. A picture of the coupling region between the HCPCF output and the PV is shown in Fig. 6b. In our previous paper [17], when using a different laser wavelength (976 nm) and PV cell (silicon-based), we found that, due to the HCPCF's small numerical aperture (typically in the order of 0.02 [22]), the end of the fiber needed to be placed at a distance of 25 mm from the PV window so to attain adequate illumination on the latter and, hence, obtain suitable electrical current intensities for the PoF application. In the present work (using a laser source emitting at 1480 nm and an InGaAs-based PV), we observed that the current intensities measured between the PV terminals mildly depended on the distance between the HCPCF output and the PV entrance. Thus, in the PoF demonstration reported in this paper, the HCPCF output end has been placed at the PV window interface (Fig. 6b). The PV, in turn, has been connected to a representative camera circuit (micro color CMOS camera). When the HCPCF delivered a laser beam with a power of 0.75 W onto the PV, a current intensity of 8.7 mA and a voltage level of 4.33 V have been measured between its terminals (here, we have chosen to limit the incident optical power to this level to avoid damaging the PV component).

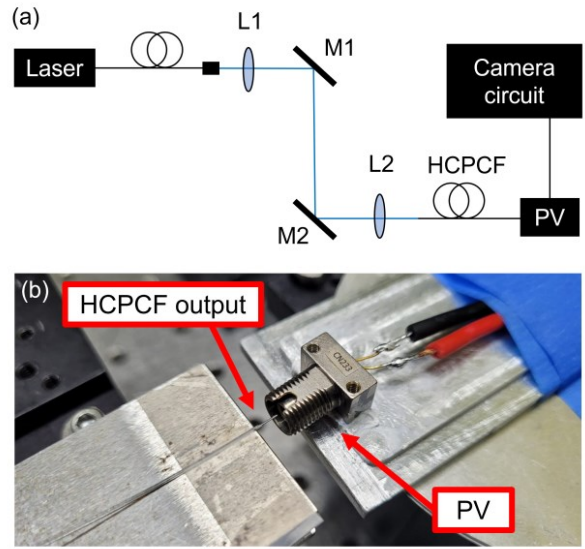

Fig. 6 (a) Diagram of the experimental setup used in the PoF demonstration (PV: photovoltaic converter). (b) Picture of the coupling region between the HCPCF output and the PV.

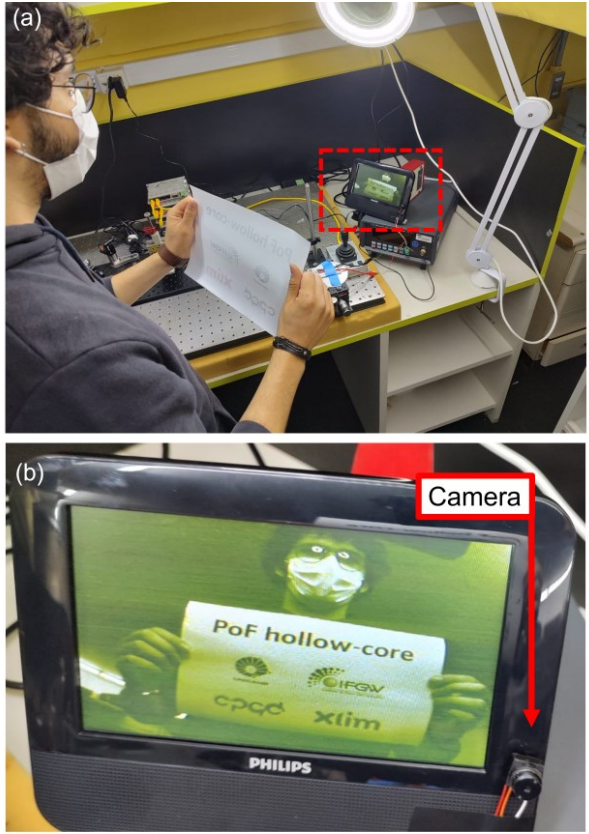

Fig. 7 (a) Picture of the HCPCF-based PoF camera system. (b) Zoom in the screen and camera sites.

The above-mentioned conditions allowed us to perform a demonstrative PoF experiment to illustrate that HCPCFs can successfully act as platforms for PoF applications. Fig. 7a shows a picture of the working camera setup, which has been powered by the HCPCF-based PoF scheme reported herein. The location of the TV screen and camera are identified by a dashed red rectangle in Fig. 7a and a zoomed picture of the camera and screen sites is provided in Fig. 7b. Fig. 7b confirms that the laser beam delivered by the HCPCF has been able to activate the tested camera circuit.

## 4. Perspectives for practical application of HCPCFs in PoF

The PoF experiment reported herein demonstrates the utilization of HCPCFs as a valid route for the development of new PoF systems. Indeed, the use of HCPCFs can be seen as a promising path for next-generation PoF schemes due to HCPCFs' capability of transporting laser beams at power levels that solid-core fiber optics cannot address. Moreover, as the endeavors of the HCPCF community have allowed achieving an impressive reduction of the HCPCF attenuation figures in the latest years, the binomial high-power transmission capability and ultralow loss stands out as the HCPCF attributes which may justify future efforts on the optimization of HCPCF schemes for PoF. It may allow, for example, the development of high-power PoF systems for outlying areas and submarine applications.

Furthermore, the modal characteristics of IC HCPCF can be tailored by adequately designing the fiber microstructure. It allows, for example, to accomplish fibers displaying an effective single-mode operation (by providing efficient filtering of the core high-order modes via resonant coupling with airy cladding modes) [5] and fibers capable of favoring the transmission of high-order modes (by suitably changing the azimuthal positioning of the cladding tubes) [23]. Additionally, IC HCPCFs exhibit low bending loss levels in the infrared, typically lower than 0.1 dB/turn for bending radii within 20 cm and 5 cm [24]. Such properties can be set by the needs of the aimed application and, hence, demonstrate the versatility of IC HCPCFs to act as building blocks for the development of new fiber devices.

Finally, as recently demonstrated in the literature, HCPCFs can be successfully integrated into current fiber optics systems. High-quality interconnections between HCPCFs and standard solid-core fibers have been accomplished by proceeding with splicing and gluing procedures [25, 26], for example. Regarding the splicing of HCPCFs with standard fiber optics, reverse tapering techniques appear as a successful approach for achieving low-loss interconnections. The latter allows for an adiabatic transference of the guided modes between the connected fibers by accomplishing adequate mode-matching conditions. Splicing loss figures as low as 0.37 dB could be attained by following the reverse-tapering process [25]. Moreover, it is worth mentioning that a 0.1 dB level splicing loss between HCPCF sections has also been reported [27]. Such conditions grant a bright framework for the integration of HCPCFs into the existing fiber optics ecosystem. Indeed, this further broadens HCPCF application possibilities and, in particular, provides a promising landscape for the employment of HCPCF in next-generation PoF systems.

## 5. Conclusions

In this manuscript, we have developed on the utilization of HCPCFs as promising platforms for next-generation PoF systems. Thus, we reported on the application of a tubular-lattice HCPCF acting as the transmission medium for delivering a continuous-wave laser beam at 1480 nm onto a photovoltaic converter. The terminals of the photovoltaic converter, in turn, have been connected to a representative camera circuit for an illustrative PoF experiment, which confirmed the feasibility of using HCPCFs within PoF scenario. We envisage that the initial PoF experiment reported herein will stimulate new developments towards novel HCPCF-based PoF schemes which, thanks to HCPCFs' outstanding capability of transporting high-power laser beams with ultralow loss, may allow elevating the current power limits of existing PoF systems and, hence, significantly impact the field of PoF applications.

## Acknowledgments

J. H. O. acknowledges Fundação de Amparo à Pesquisa do Estado de São Paulo (FAPESP) for financial support (grant 2021/13097-9). C. M. B. C. and J. B. R. thank Conselho Nacional de Desenvolvimento Científico e Tecnológico (CNPq), grants 309989/2021-3 and 303791/2021-7 respectively. This project has been developed within Sisfoton-MCTI's Laboratories cooperation.

REFERENCES

1. T. A. Birks, P. J. Roberts, P. S. J. Russell, D. M. Atkin, T. J. Shepherd, "Full 2-D photonic bandgaps in silica/air structures," Electronics Letters 31, 1941-1943 (1995).
2. F. Couny, F. Benabid, P. J. Roberts, P. S. Light, M. G. Raymer, "Generation and photonic guidance of multi-octave optical frequency combs," Science 318, 1118-1121 (2007).

3.  Y. Wang, F. Couny, P. J. Roberts, F. Benabid, "Low loss broadband transmission in hypocycloid-core Kagome hollow-core photonic crystal fiber," Optics Letters 36, 669-671 (2011).

4.  J. H. Osório, F. Amrani, F. Delahaye, A. Dhaybi, K. Vasko, G. Tessier, F. Giovanardi, L. Vincetti, B. Debord, F. Gérôme, F. Benabid, "Hollow-core fibers with ultralow loss in the ultraviolet range and sub-thermodynamic equilibrium surface-roughness," Conference on Lasers and Electro-Optics, (CLEO, 2022), paper SW4K.6.

5.  F. Amrani, J. H. Osório, F. Delahaye, F. Giovanardi, L. Vincetti, B. Debord, F. Gérôme, F. Benabid, "Low-loss single-mode hybrid-lattice hollow-core photonic-crystal fibre," Light: Science and Applications 10, 7 (2021).

6.  G. T. Jasion, H. Sakr, J. R. Hayes, S. R. Sandoghchi, L. Hooper, E. N. Fokoua, A. Saljoghei, H. C. Mulvad, M. Alonso, A. Taranta, T. D. Bradley, I. A. Davidson, Y. Chen, D. J. Richardson, F. Poletti, "0.174 dB/km hollow core double nested antiresonant nodeless fiber (DNANF)," 2022 Optical Fiber Communications Conference and Exhibition (OFC), 2022, 1-3.

7.  M. Chafer, J. H. Osório, A. Dhaybi, F. Ravetta, F. Amrani, F. Delahaye, B. Debord, C. Cailteau-Fischbach, G. Ancellet, F. Gérôme, F. Benabid, "Near- and middle-ultraviolet reconfigurable Raman source using a record-low UV/visible transmission loss inhibited-coupling hollow-core fiber," Optics & Laser Technology 147, 107678 (2022).

8.  P. Poggiolini, F. Poletti, "Opportunities and challenges for long-distance transmission in hollow-core fibres," Journal of Lightwave Technology 40, 6, 1605-1616 (2022).

9.  B. Debord, M. Alharbi, T. Bradley, C. Fourcade-Dutin, Y. Y. Wang, L. Vincetti, F. Gérôme, F. Benabid, "Hypocycloid-shaped hollow-core photonic crystal fiber Part I: arc curvature effect on confinement loss," Optics Express 21, 28597-28608 (2013).

10. B. Debord, F. Gérôme, P. Paul, A. Husakou, F. Benabid, "2.6 mJ energy and 81 GW peak power femtosecond laser-pulse delivery and spectral broadening in inhibited coupling Kagome fiber," in CLEO:2015, OSA Technical Digest (online) (Optica Publishing Group, 2015), paper STh4L.7.

11. S. Hädrich, J. Rothhardt, S. Demmler, M. Tschernajew, A. Hoffmann, M. Krebs, A. Liem, O. de Vries, M. Plötner, S. Fabian, T. Schreiber, J. Limpert, A. Tünnermann, "Scalability of components for kW-level average power few-cycle lasers," Applied Optics 55, 1636-1640 (2016).

12. H. C. H. Mulvad, S. Abokhamis Mousavi, V. Zuba, L. Xu, H. Sakr, T. D. Bradley, J. R. Hayes, G. T. Jasion, E. Numkam Fokoua, A. Taranta, S. U. Alam, D. J. Richardson, F. Poletti, "Kilowatt-average-power single-mode laser light transmission over kilometre-scale hollow-core fibre," Nature Photonics 16, 448-453 (2022).

13. J. B. Rosolem, "Power-over-fiber applications for telecommunications and for electric utilities," IntechOpen, 10.5772/68088 (2017).

14. K. Seo, N. Nishimura, M. Shiino, R. Yugushi, H. Sasaki, "Evaluation of high-power endurance in optical fiber links," Furukawa Review, 24:17-22 (2003).

15. J. Li, A. Zhang, G. Zhou, J. Liu, C. Xia, Z. Hou, "A large-core microstructure optical fiber for co-transmission of signal and power," Journal of Lightwave Technology 39, 4511-4516 (2021).

16. M. Matsuura, N. Tajima, H. Nomoto, D. Kamiyama, "150-W power-over-fiber using double-clad fibers," Journal of Lightwave Technology 38, 401-408 (2020).

17. J. H. Osório, J. B. Rosolem, F. R. Bassan, F. Amrani, F. Gérôme, F. Benabid, C. M. B. Cordeiro, "Hollow-core photonic crystal fibers as platforms for power-over-fiber applications," 4th Optical Wireless and Fiber Power Transmission Conference (OWPT), 2022.

18. B. Debord, A. Amsanpally, M. Chafer, A. Baz, M. Maurel, J. M. Blondy, E. Hugonnot, F. Scol, L. Vincetti, F. Gérôme, F. Benabid, "Ultralow transmission loss in inhibited-coupling guiding hollow fibers," Optica 4, 209-217 (2017).

19. L. Vincetti, "Empirical formulas for calculating loss in hollow core tube lattice fibers," Optics Express 24, 10313-10325 (2016).

20. Broadcom Application Notes, "Optical Power Components Optimizing Optical Power Converter Output," Available in https://docs.broadcom.com/doc/AFBR-POCxxxL-AN

21. T. Shan, X. Qi, "Performance of series connected GaAs photovoltaic converters under multimode optical fiber illumination," Advances in Optoelectronics, 824181 (2014).

22. F. Delahaye, F. Gérôme, F. Amrani, A. Uterhuber, K. Vasko, B. Debord, M. Andreana, F. Benabid, "Double-clad hollow-core photonic crystal fiber for nonlinear optical imaging," in Conference on Lasers and Electro-Optics, paper AF2Q.2 (2017).

23. J. H. Osório, M. Chafer, B. Debord, F. Giovanardi, M. Cordier, M. Maurel, F. Delahaye, F. Amrani, L. Vincetti, F. Gérôme, F. Benabid, "Tailoring modal properties of inhibited-coupling guiding fibers by cladding modification," Scientific Reports 9, 1376 (2019).

24. M. Chafer, J. H. Osório, F. Amrani, F. Delahaye, M. Maurel, B. Debord, F. Gérôme, F. Benabid, "1-km hollow-core fiber with loss at the silica Rayleigh limit in the green spectral region," IEEE Photonics Technology Letters 31, 9 (2019).

25. C. Wang, R. Yu, B. Debord, F. Gérôme, F. Benabid, K. S. Chiang, L. Xiao, "Ultralow-loss fusion splicing between negative curvature hollow-core fibers and conventional SMFs with a reverse-tapering method," Optics Express 29, 22470-22478 (2021).

26. D. Suslov, M. Komanec, E. R. Numkam Fokoua, D. Dousek, A. Zhong, S. Zvánovec, T. D. Bradley, F. Poletti, D. J. Richardson, R. Slavík, "Low loss and high performance interconnection between standard single-mode fiber and antiresonant hollow-core fiber," Scientific Reports 11, 8799 (2021).

27. Y. Hong, H. Sakr, N. Taengnoi, K. R. H. Bottrill, T. D. Bradley, J. R. Hayes, G. T. Jasion, H. Kim, N. K. Thipparapu, Y. Wang, A. A. Umnikov, J. K. Sahu, F. Poletti, P. Petropoulos, D. J. Richardson, "Multi-band direct-detetion transmission over an ultrawide bandwidth hollow-core NANF," Journal of Lightwave Technology 38, 2849-2857 (2020).